\documentclass[12pt]{article}
\usepackage{amsmath}
\begin{document}
%
\title{\bf Physics of Complex Magnetic Materials: Quasiparticle Many-Body Dynamics  }
%
\author{
A. L. Kuzemsky \thanks {E-mail:kuzemsky@theor.jinr.ru;
http://theor.jinr.ru/~kuzemsky}
\\
{\it Bogoliubov Laboratory of Theoretical Physics,} \\
{\it  Joint Institute for Nuclear Research,}\\
{\it 141980 Dubna, Moscow Region, Russia.}}
\date{}
\maketitle
\begin{abstract}
A brief survey of the author's works on the quantum theory of
magnetism.
Theoretical foundation and applications of the generalized
spin-fermion ( sp-d ) exchange lattice model to various magnetic
systems, e.g., rare-earth metals and compounds,  and magnetic and
diluted magnetic semiconductors are discussed briefly. The main
emphasis is put on the dynamic behavior of two interacting
subsystems, the localized spins and spin density of itinerant
carriers. A nonperturbative many-body approach
was used to describe the quasiparticle dynamics. 
The approach permits to investigate and
clarify  the role of various interactions and disorder effects in
combined quantum models of magnetism in unified and coherent fashion.   
\end{abstract}
%

The existence and properties of localized and itinerant magnetism
in metals, oxides, and alloys, and their interplay is an
interesting but not yet fully understood problem of quantum
theory of magnetism\cite{tyab,ku00}.   The behavior and
the true nature of the electronic and spin states, and their
quasiparticle dynamics are of central importance to the
understanding of physics of correlated systems such as magnetism
and Mott-Hubbard metal-insulator transition in metals and oxides,
magnetism and heavy fermions  in rare-earths compounds, and
anomalous transport properties in perovskite manganites. This
class of systems is characterized by  complex, many-branch
spectra of elementary excitations. Moreover, the correlation
effects ( competition and interplay of Coulomb correlation,
direct or indirect exchange, sp-d hybridization, electron-phonon
interaction, disorder, etc.) are essential\cite{ku97,kuz02}. These
materials are systems of great interest both intrinsically and as
a possible source of understanding the magnetism of matter
generally~\cite{ku02,kuz99}. 
There has been considerable interest in identifying the
microscopic origin of quasiparticle states~\cite{ku97,kuz02,kuz99} in these systems and a
few  model approaches have been proposed. Many magnetic and
electronic properties of rare-earth metals and
compounds, and magnetic semiconductors~\cite{nag}
and related  complex magnetic materials may  reasonably  be interpreted  in terms
of combined spin-fermion models  which include interacting
spin and charge subsystems~\cite{ku99,ku04,kuz04}. This
approach permits one to describe  significant and interesting
physics, e.g.,  bound states and magnetic
polarons~\cite{ku04}, anomalous transport properties, etc. \\
The problem of   adequate physical description  within various
types of  spin-fermion model has  intensively been
studied during the last decades, especially in the context of
magnetic and transport properties of rare-earth and transition
metals and their compounds and magnetic semiconductors.\\
Substances which we refer   to as magnetic semiconductors,
occupy  an intermediate position between magnetic metals and
magnetic dielectrics. Magnetic semiconductors are characterized
by the existence of two well defined subsystems, the system of
magnetic moments which are localized at lattice sites, and a band
of itinerant or conduction carriers (conduction electrons or
holes). Typical examples are the Eu-chalcogenides, where the local
moments arise  from 4f electrons of the Eu ion, and the spinell
chalcogenides containing $Cr^{3+}$ as a magnetic ion~\cite{nag}. There is
experimental evidence of a substantial mutual influence of spin
and charge subsystems in these compounds. This is possible due to
the $sp-d(f)$ exchange interaction of the localized spins and
itinerant charge carriers.  
More recent efforts have been directed to the study of the
properties of  diluted magnetic
semiconductors (DMS)\cite{kuz04}. 
Further attempts have been
made to study and exploit carriers which are exchange-coupled to
the localized spins.
The effect of carriers on the magnetic ordering temperature is  
found to be very strong in  
 DMS . Diluted magnetic semiconductors are
mixed crystals in which magnetic ions (usually $Mn^{++}$) are
incorporated in a substitutional position of the host ( typically
a II-VI or III-V ) crystal lattice. The diluted magnetic
semiconductors offer a unique possibility for a gradual change of
the magnitude and sign of exchange interaction by means of
technological control of carrier concentration and band parameters.
This field is
very active and there are many aspects to the problem. A lot of
materials were synthesized and tested. The
new material design approach to fabrication of new functional
diluted magnetic semiconductors resulted in  producing a
variety of compounds . The presence of
the spin degree of freedom in DMS may lead to a new semiconductor
spin electronics which will combine the advantages of the
semiconducting devices with the new features due to the
possibilities of controlling the magnetic state.
However, the coexistence of ferromagnetism and semiconducting
properties in these compounds require a suitable theoretical
model which would describe well both the magnetic cooperative
behavior and the semiconducting properties as well as a rich field
of interplay between them. The majority of theoretical papers on
DMS studied their properties mainly within the mean field
approximation and continuous media terms. In  a picture like this
the disorder effects, which play an essential
role, can be taken into account roughly
only. Moreover, there are different opinions on the intrinsic
origin and the nature of disorder in DMS. 
Thus, many experimental and
theoretical investigations call for a better understanding of the
relevant physics and  the   nature of solutions ( especially
magnetic ) within  the lattice spin-fermion
model\cite{ku99,ku04,kuz04}. 
For treating the problems we used a nonperturbative many-body approach, the formalism of the method of
Irreducible Green Functions (IGF)~\cite{kuz02}. This IGF method
allows one to describe  quasiparticle spectra with damping for
many-particle systems on a lattice with complex spectra and a
strong correlation in a very general and natural way. This scheme
differs from the traditional method of decoupling of an infinite
chain of  equations~\cite{tyab} and permits a construction of the
relevant dynamic solutions in a self-consistent way at the level
of the Dyson equation without decoupling the chain of equations
of motion for the GFs.\\ 
In paper~\cite{ku04}
the concepts of bound and scattering states were analysed and developed to
elucidate the nature of itinerant charge carrier states in
magnetic semiconductors and similar complex magnetic materials.
By contrasting the scattering and bound states of carriers within
the $s-d$ exchange model, the nature of bound states at finite
temperatures was clarified. The free magnetic polaron at certain
conditions is realized as a bound state of the carrier (electron
or hole) with the spin wave. Quite generally, a self-consistent
theory of a magnetic polaron was formulated within  
the IGF   method  which was used to describe the
quasiparticle many-body dynamics at finite temperatures. Within
the above  approach we elaborated a self-consistent
picture of   dynamic behavior of two interacting subsystems, the
localized spins and the itinerant charge carriers. In particular,
it was shown that the relevant generalized mean fields emerges
naturally within our formalism. At the same time, the correct
separation of elastic scattering corrections permitted one to
consider the damping effects (inelastic scattering corrections)
self-consistently. The damping of magnetic
polaron state, which is quite different from the damping of the
scattering states, finds a natural interpretation within the
present self-consistent scheme. \\
In  paper~\cite{kuz04}, we applied the IGF
formalism to consider  quasiparticle spectra for the lattice
spin-fermion model consisting of two interacting subsystems. It
was the purpose of that paper to explore more fully the notion of
Generalized Mean Fields (GMF)~\cite{kuz02} which may arise in the
system of interacting localized spins ( including effects of
disorder ) and lattice fermions to justify and understand the
nature of the relevant mean fields. Background and applications
of the generalized spin-fermion ( sp-d ) exchange model to
magnetic and diluted magnetic semiconductors were discussed in
some detail. The capabilities of the model to describe
quasiparticle spectra were investigated. The key problem of most
of the work was the formation of spin excitation spectra under
various conditions on the parameters of the model. In paper~\cite{kuz04}, we  concentrated on
the description of the magnetic excitation spectra and treated the
disorder effects in the simplest VCA (virtual crystal approximation) to emphasize  
the need for a suitable definition of the
relevant generalized mean fields and for internal
self-consistency in the description
of the spin quasiparticle many-body dynamics.\\
 Thus, we were able to calculate the
quasiparticle spectra and GMF of the  magnetic semiconductors
consisting of two interacting charge and spin subsystem  within
the lattice spin-fermion model 
to analyze the role and influence of the Coulomb correlation,
exchange, and effects of disorder in a unified and coherent fashion.

%

\end{document}